\documentclass{iopjournal}

\usepackage{hyperref}
\hypersetup{
	colorlinks=true,
	linkcolor=blue,
	urlcolor=blue,
	citecolor=blue} 

\usepackage{amssymb}
\usepackage{amsmath}
\usepackage{csquotes}

\begin{document}

\title{Unifying pendulum-like dynamical regimes via complex time}

\author{Teepanis Chachiyo}
\email{teepanisc@nu.ac.th}
\affil{Department of Physics, Faculty of Science, Naresuan University, Phitsanulok 65000, Thailand}

\begin{abstract}
We present an alternative derivation of exact solutions for pendulum-like systems across all dynamical regimes using a standard technique in an undergraduate mathematical physics course: the Cauchy residue theorem, entirely independent from the traditional Jacobi elliptic functions. The solutions are exact in both the time and frequency domains, providing continuous trajectories and precise frequency decompositions. Pendulum-like dynamics is a foundational model across many areas of physics, underlying systems ranging from classical nonlinear oscillators to superconducting qubits and cold-atom tunneling platforms. While its time-domain solutions are well-known in terms of Jacobi elliptic functions, its obscured frequency-domain solutions have led a large body of theoretical and experimental work to develop and apply approximate methods when studying its spectral behavior. We discover that all regimes arise from a single spectral kernel, with parity selection distinguishing the periodic motions and the separatrix representing their discrete-to-continuum limit. Regime changes thus correspond to symmetry-driven reorganizations in frequency space rather than changes in the underlying spectral structure, with the stopping trajectory representing the continuum limit. By completely bypassing the traditional Jacobi elliptic functions, and relying on the symmetrical structure of the complex time plane, the derivation shows that the symmetrical spectral structure is not merely an algebraic artifacts of elliptic integrals, but fundamentally connected to the dynamical symmetries of the system. The derivation presented here not only serves as a powerful pedagogical exercise for this long-standing physics problem, but also reveals a highly symmetrical spectral organization in nonlinear dynamics.
\end{abstract}


\section{Introduction}

\begin{figure}[t]
	\begin{center}
		\includegraphics[width=0.6\textwidth]{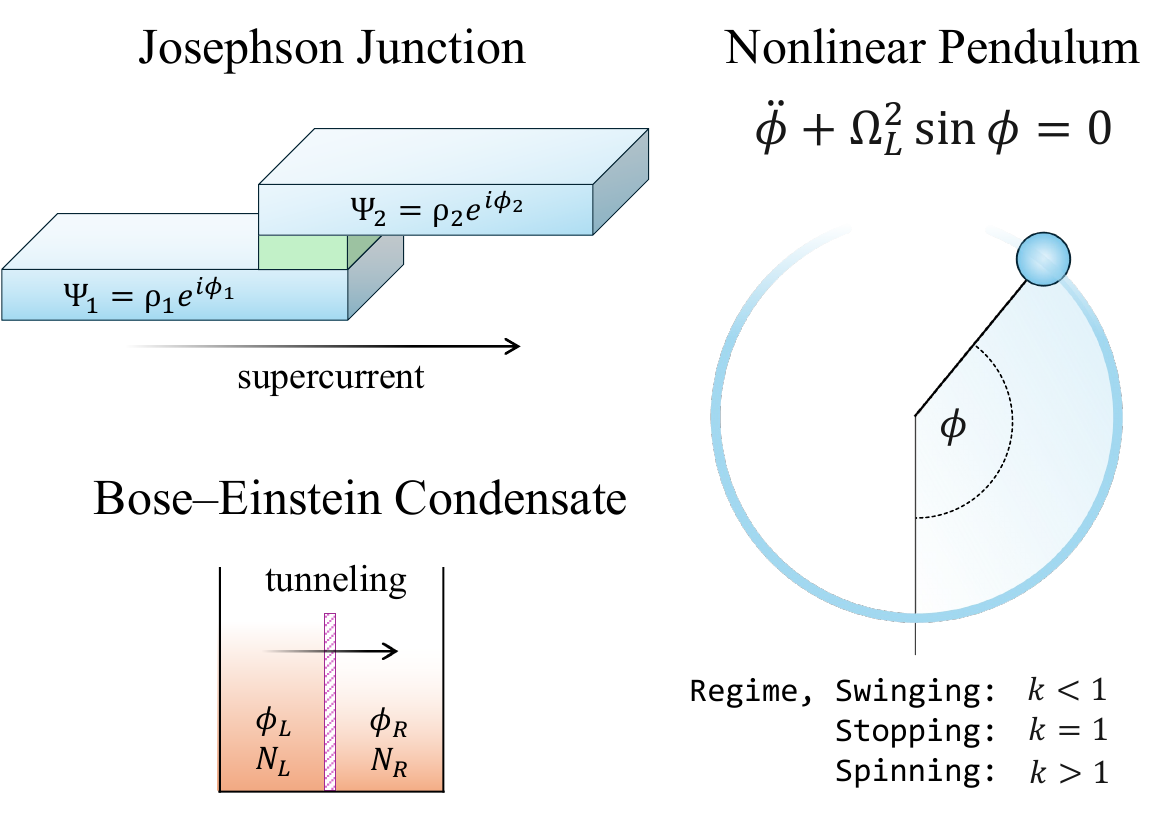}
		\caption{\label{fig_pendulum-like} \textbf{Pendulum-like dynamics across classical and quantum systems.}
			(a) Superconducting Josephson junction, where the phase difference 
			$\phi = \phi_2 - \phi_1$ governs the supercurrent.
			(b) Neutral-atom (bosonic) Josephson junction with populations 
			$N_L$ and $N_R$ and tunneling between condensates.
			(c) Nonlinear pendulum with angular displacement $\phi$.
			All systems are described by the nonlinear phase equation 
			$\ddot{\phi} + \Omega_L^2 \sin\phi = 0$, which admits
			swinging, stopping, and spinning regimes.
		}
	\end{center}
\end{figure}

Pendulum-like dynamics, as shown in Fig.~\ref{fig_pendulum-like}, remains an active topic of research \cite{Moatimid2022,Hinrichsen2023a,Hinrichsen2023b,Zhang2024,Kontomaris2024}, with broad applications ranging from superconducting Josephson junctions \cite{MacDonald1983,Mangin2016} and Bose--Einstein condensate tunneling \cite{Marino1999,Pigneur2018} to energy harvesting and artificial intelligence \cite{Marszal2017,Fu2023,Feng2024,Krishnapriyan2023}. For example, pendulum-based mechanisms can harvest energy from human motion \cite{Fu2023} or ocean waves via piezoelectric coupling \cite{Feng2024}, while machine-learning models employ analytical pendulum solutions to recognize continuous nonlinear dynamics \cite{Krishnapriyan2023}. 

Since the groundbreaking study by Galileo on the pendulum in the 1600s~\cite{Matthews2004}, its exact spectral behavior has remained elusive. This is evident from the following published works. First, recent experimental studies on the nonlinear pendulum ~\cite{Gil2008,Hinrichsen2020} still used approximate Fourier coefficients obtained from previous theoretical work on perturbation methods~\cite{Fulcher1976,Zilio1982}.  Second, recent theoretical studies applied an approximation scheme to find the Fourier coefficients and validated their results through numerical tests~\cite{Mndez2010}. Third, numerical methods are still being used to perform Fourier analysis of the nonlinear pendulum in both recent and early work~\cite{Singh2018,Simon1979}. Fourth, the traditional time-domain solution involves a Jacobi elliptic sine function, $\text{sn}(z, k)$. It is true that an exact Fourier series of $\text{sn}(z, k)$ is known~\cite[{22.11.E1}]{NIST:DLMF}. However, the form of the exact $\phi(t)$ is such that the elliptic sine is embedded inside an arcsine function, $\phi(t) = 2 \arcsin[k\,\text{sn}(\Omega_L t,\,k)]$~\cite{Lima2018,Belndez2007}. Unless a clever mathematical manipulation is performed to isolate it, knowing the exact Fourier series of $\text{sn}(z, k)$ does not constitute an exact Fourier series of $\phi(t)$.

Evidently, a large body of theoretical and experimental work from the early 1970s to the 2020s has developed and applied approximate methods when studying the spectral behavior of the pendulum. This suggests that the exact frequency-domain solution across all dynamical regimes has been obscured, perhaps due to the inherent complexity of the Jacobi elliptic framework.

Here, we present an alternative derivation of exact solutions for pendulum-like systems across all dynamical regimes using a standard technique in an undergraduate mathematical physics course: the Cauchy residue theorem, entirely independent from the traditional Jacobi elliptic functions. The solutions are exact in both the time and frequency domains, providing trajectories and precise frequency decompositions. 

Before presenting the derivation, we elaborate on the educational focus in this work, explaining the intended student levels, teaching context, and specific contribution to students' understanding. Pendulum motion has always been an important part of physics education, ranging from the freshman introductory level and sophomore intermediate level to the famous MIT demonstration by Walter Lewin~\cite{Lewin2011Love}. Strategically placed throughout this paper and its online supplementary material, the presentation in this work is designed to foster students' understanding of multiple levels in the undergraduate physics curriculum. 

At the introductory level~\cite{Serway2016Physics,Halliday2013Fundamentals}, a typical freshman textbook often presents the pendulum as simple harmonic motion, making an approximation that the amplitude is small, typically less than 5 degrees. This limitation is an opportunity for students to ask what happens if the pendulum is allowed to swing at a larger amplitude. The online supplementary material in this work (see Data Availability) provides an intuitive answer to this question. A simulation is presented to give students a real-time understanding of how the pendulum motion is classified into three regimes. Python code is given so that students can change the parameters themselves and learn exactly how the frequency-domain decompositions are implemented and turned into smooth time-domain trajectories.

While it is unrealistic for freshmen to follow the complex time derivation, an instructor can focus on the conceptual anchors that a motion can be nonlinear and exhibit multiple dynamical regimes. Thus, the familiar pendulum is used as a steppingstone toward a deeper understanding that students will master in the later courses. We believe that a good lesson should not only be comprehensive within its designed scope, but also hint to a deeper exploration beyond the horizon.

Unlike the Jacobi elliptic functions, the exact results in this work are given in terms of elementary functions, save for one: the complete elliptic integral of the first kind $K(k)$. However, this special function is not a computational barrier for designing an advanced pendulum laboratory accessible to introductory-level students. Our previous work on the accurate approximation for the full range of modulus, $K \approx \frac{1}{n}\ln [(\frac{4}{\sqrt{1-k^2}})^n+b]$ with constants $n$ and $b$ defined in Ref. ~\cite{Chachiyo2026}, provides an accessible computational tool for students to plot its analytical solution as compared to the observed trajectory from Tracker software~\cite{TrackerSoftware2026}. Being valid for the full range, $0 < k < 1$, is also necessary because the analytical solution requires both $K(k)$ and $K(\sqrt{1-k^2})$, meaning when $k \approx 0$ then $\sqrt{1-k^2} \approx 1$. Its design guarantees that the approximation is suitable for nonlinear pendulum motion. 

For example, a simple adaptation of the freshman physics laboratory course would be to allow the pendulum to swing at an arbitrary amplitude $\theta_0$. The approximation directly gives the period $T \approx T_0 \frac{2}{n \pi} \ln[ (\frac{4}{\cos (\theta_0/2)})^n+ b]$ where $T_0$ is the simple harmonic period. The expression also allows for its inverse to be easily evaluated. Students can inversely predict the amplitude $\theta_0$ for a given nonlinear period $T$, a feat that would otherwise require heavy computational root finding method to achieve.

At the intermediate level~\cite{Marion1995,Tong2025Classical,Tong2025Quantum}, there are a number of core physics courses that can incorporate the complex time derivation in this work into their material. Directly related is a mathematical methods in physics course, which typically includes complex analysis and special functions. A classical mechanics course can introduce the concept of dynamical regimes where the dynamics appears drastically different merely by adjusting physical parameters. A typical example is a damped mass-spring system where the motion can be classified into underdamped, critically damped, and overdamped regimes. Here, the three regimes of pendulum motion can also be introduced with a clear mental hook while maintaining precise mathematical rigor. A quantum mechanics course often deviates entirely from the classical ones, presenting a new set of quantum postulates and analysis.  Section \ref{sec:quantum_analogue}: Pedagogical Overview of Quantum Analogues, is a suitable addition because it creates a narrative arc. It explains how a classical pendulum can indeed be used to study the dynamics of state-of-the-art quantum research such as the superconducting Josephson junction and Bose--Einstein condensate tunneling.

Taken together, the derivation presented here not only serves as a pedagogical exercise for this centuries-old physics problem, but also reveals a highly symmetrical spectral organization in nonlinear dynamics.

\section{Preliminary}

Energy-conserving pendulum-like dynamics are governed by the following equation:

\begin{equation}
	\frac{d^2}{dt^2}\phi(t) + \Omega_L^2 \sin \phi = 0. \label{eq_pendulum_eq}
\end{equation}

Here, $\phi(t)$ and $\Omega_L$ are the characteristic dynamical phase and the angular frequency of \emph{linear} oscillation, respectively. $\Omega^2_L$  depends on physical parameters of the system. For example, $\Omega^2_L  = {g/L}$ in a classical pendulum~\cite{Marion1995,Taylor2004},  ${2e I_c}/({\hbar C})$ in an ideal Josephson junction~\cite{Roth2023}, or $[\frac{2 J}{\hbar}\sqrt{\Lambda + \lambda}]^2$ in Bose--Einstein condensate tunneling~\cite{Pigneur2018}.

Nonlinear motions of an energy-conserving pendulum can be divided into three regimes: Swinging, Stopping, and Spinning. These names are coined in this work to enhance intuitive interpretation. Consider the case where the pendulum starts at the bottom, $\phi_0 = 0$, with only $\omega_0$ to initiate the motion. Due to conservation of energy, the angular speed at the bottom is always maximal; thus, $\omega_m$ will be used to denote the angular speed at the bottom.

If $\omega_m$ that initiates the motion is too low, the pendulum will not reach the top and will inevitably fall back down. This is referred to as \enquote{Swinging} motion. If $\omega_m$ is just sufficient for the pendulum to reach the top, it will approach the top but never quite get there. This is because conservation of energy dictates that the closer the pendulum gets to the top, the slower it moves. Ideally and conceptually, the pendulum approaches the top asymptotically, and the angular speed decreases exponentially in time. This is referred to as \enquote{Stopping} motion. Finally, if $\omega_m$ is too large, the pendulum will continue to rotate in one direction. This is referred to as \enquote{Spinning} motion.

There exists a critical value of angular speed $\omega_c$, which separates the motion into the three distinct regimes. Using conservation of energy, $E = \frac{1}{2} \dot{\phi}^2 - \Omega_L^2 \cos \phi$, the maximal angular speed, which occurs at $\phi = 0$, that provides just enough kinetic energy to reach the top at $\phi = \pi$ is:

\begin{equation}
		\omega_c = 2 \Omega_L.
\end{equation}

Maximal speed $\omega_m$ that is lower, equal, or greater than $\omega_c$ will result in the swinging, stopping, and spinning regimes, respectively. We define the modulus $k$ as the parameter that continuously traverses across all dynamical regimes.

\begin{equation}
k \equiv \frac{\omega_m}{\omega_c} \quad \begin{cases} k < 1 \quad \text{swinging} \\ k = 1 \quad \text{stopping} \\ k > 1 \quad \text{spinning} \end{cases}
\end{equation}

\section{Method}

For simplicity and without loss of generality, we start with the time-normalized ODE:

\begin{equation}
	\ddot{\phi} + \sin \phi = 0,
\end{equation}
 where $\Omega_L= 1$ and $k = \omega_m/2$. The time scaling $t \to t/\Omega_L$ can always be reintroduced afterward.
 
 We shift the focus to angular velocity $\omega(t) = \dot{\phi}(t)$. Unlike the angular displacement $\phi(t)$ which can increase without bound for the spinning motion, $\omega(t)$ is strictly periodic due to energy conservation. The angular velocity can be expressed in a standard complex Fourier series:

\begin{equation}
	\omega(t) = \sum_{n = -\infty}^{+\infty} w_n e^{+i n \Omega_0 t}.
\end{equation}

Here, $\Omega_0 = 2 \pi /T $ is the angular frequency.   By orthogonality, the coefficient $w_n$ is extracted by integrating over one period $T$.

\begin{equation}
w_n = \frac{1}{T} \int_0^T \omega(t) e^{- i n \Omega_0 t} dt.
\end{equation}

The aim is to find the exact $w_n$ for both swinging and spinning regimes, without explicitly knowing the form of $\omega(t )$ itself. The key is to promote the time $t$ to the complex plane, and apply the Cauchy residue theorem to evaluate $w_n$:

\begin{equation}
\oint_C \frac{1}{T} \omega(t) e ^{- i n \Omega_0 t} dt = 2 \pi i \sum \text{Res}. \label{eq:Cauchy}
\end{equation}

However, before applying the Cauchy residue theorem, there are a number of quantities such as the period, the poles (singularities), the residues, and the complex contours that need to be made explicit.

\subsection{The Period}

\begin{figure}
	\begin{center}
		\includegraphics[width=0.75\textwidth]{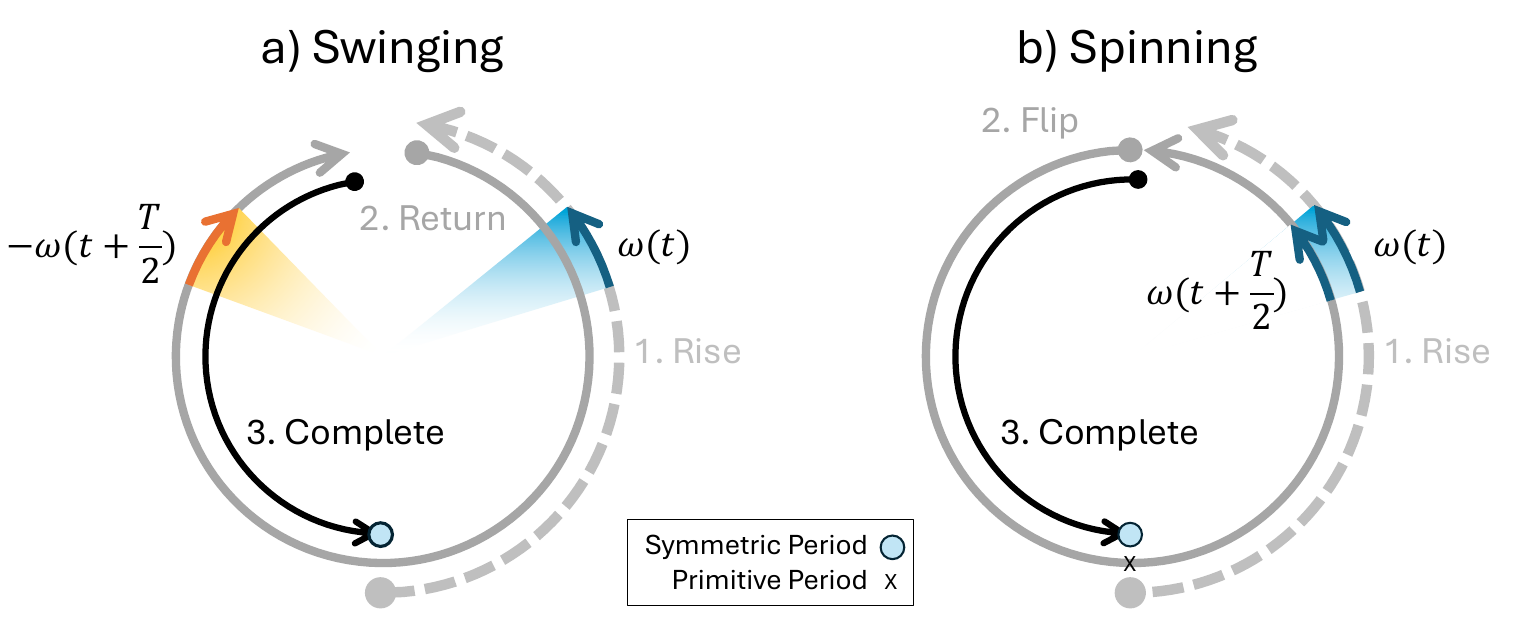}
		\caption{\label{fig_periods} \textbf{Primitive and symmetric periods of swinging and spinning regimes.}
			(a) The pendulum rises, returns, and completes its revolution.
			(b) The pendulum rises, flips, and completes its revolution. As indicated by the cross symbol, half of the flipping motion brings it to the primitive period.
			However, if we extend the flipping motion all the way to the top, followed by the completing motion, the spinning period becomes symmetrical to the swinging regime, $T = 4 K$.
		}
	\end{center}
\end{figure}

The expressions for the primitive periods for the swinging and spinning regimes are well-known in standard textbooks~\cite{Marion1995,Ochs2011}. We will not repeat the derivation here in the interest of keeping the focus on the Fourier coefficients.

\begin{equation}
\begin{aligned}
&  \text{swinging: } && \quad T_\text{prim} = 4 K(k), & \; k < 1, \\
& \text{spinning: } && \quad T_\text{prim} = 2 K(1/k)/k, & \; k > 1.
\end{aligned}
\end{equation}

Although $K(k)$ becomes complex as $k > 1$, its real part is $K(1/k)/k$ which is precisely proportional to the primitive spinning period. Therefore, the periods can be written more uniformly as:

\begin{equation}
	\begin{aligned}
		&  \text{swinging: } && \quad T_\text{prim} = 4 \, \text{Re}[K(k)], & \; k < 1, \\
		& \text{spinning: } && \quad T_\text{prim} = 2 \, \text{Re}[ K(k)], & \; k > 1.
	\end{aligned}
\end{equation}

The key perspective shift that allows spectral unification across all dynamical regimes is to realize that the symmetric period of the spinning regime is actually double its primitive period which is $T = 4 \, \text{Re}[K(k)]$. Mathematically, doubling a primitive period is an admissible choice, but more fundamentally is the physical argument behind it. 

Figure \ref{fig_periods} illustrates this concept. Instead of terminating the flipping motion at its primitive period (e.g., denoted $\times$), we extend the flipping motion further. We then see that the flipping motion is identical to the returning motion, except in reverse. This doubling of the primitive period, as shall be evident in the subsequent derivation, unlocks symmetric/anti-symmetric relations between the two regimes: the pole locations, the residues, and the odd-even harmonic sampling.

Therefore, we propose that period of all dynamical regimes takes a single symmetric form:

\begin{equation}
	T = 4 \, \text{Re}[K(k)], \; \text{abbreviated as } 4 K.
\end{equation}

The abbreviation helps keep the algebraic presentation readable. 

Having defined the universal period $T$, there are also symmetric properties of angular velocity. As shown in Fig.~\ref{fig_periods}, if time is shifted by half a period ($T/2$), the pendulum is at the exact same height but moving in the opposite direction in the swinging regime, and in the same direction in the spinning regime. Mathematically, this means that $\dot{\phi}$ is either anti-periodic or strictly periodic over a half-period.

\begin{align}
 \text{swinging:} & \quad \dot{\phi}(t) = -\dot{\phi}(t + \frac{T}{2}), \label{eq:halfperiod_swing}\\
\text{spinning:} & \quad  \dot{\phi}(t) = +\dot{\phi}(t + \frac{T}{2}).\label{eq:halfperiod_spin} 
\end{align}

\subsection{The Poles}

\begin{figure}
	\begin{center}
		\includegraphics[width=0.75\textwidth]{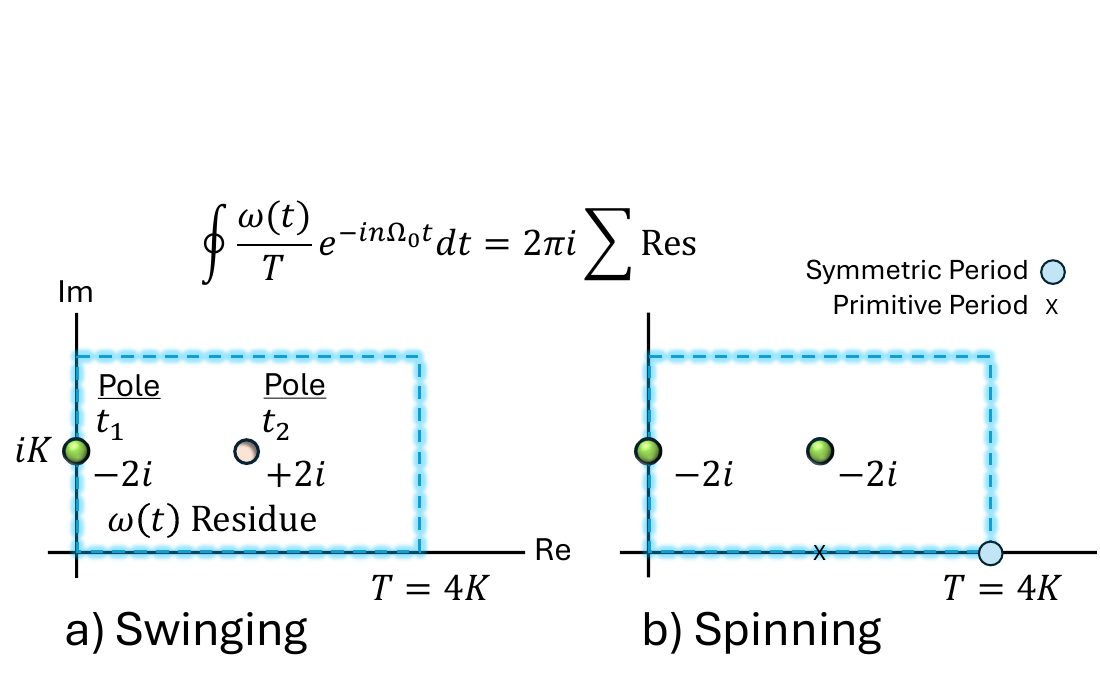}
		\caption{\label{fig_contour} \textbf{Poles, residues and contour paths.}
			(a) The residues are opposite phase.
			(b) The residues are in-phase.
			The phase difference follows directly from the anti or strictly periodicity of the angular velocity, which in turn is a result of the universal period $T = 4 K$.
		}
	\end{center}
\end{figure}

Poles are points in the complex time plane where the solution $\phi(t)$ diverges to infinity. Since the solutions are periodic, we are only interested in the poles located inside the fundamental real-time strip $0 \le \text{Re}(t) < 4 K$. It is algebraically simpler to look for poles in the velocity $\dot{\phi}(t)$. First, we establish the energy integral:

\begin{equation}
	\frac{1}{2}\dot{\phi}^2 - \cos \phi = E,
\end{equation}
where $E = 2 k^2 - 1$ across all regimes. The singularity in $\dot{\phi}$ occurs when $\cos \phi \to \infty$ which is consistent with $\phi \to +i \infty$ as we ascend into the positive imaginary-time direction. The time it takes to reach this singularity from the origin is purely imaginary. Integrating $t(\phi)$ from $\phi=0$ to $+i\infty$ yields:

\begin{equation}
	t_1 = \int_0^{+i\infty} \frac{d\phi}{\sqrt{2 E + 2 \cos \phi}} = \int_0^{+i \infty} \frac{d\phi}{2 \sqrt{k^2 - \sin^2{\phi/2}}}.
\end{equation}

To standardize this integral, we use the substitution $\sin(\phi/2) =  i k \tan \theta$. Differentiating gives $\frac{1}{2} \cos (\phi/2) d\phi =  i k \sec^2 \theta d\theta$, which transforms the integral into:

\begin{equation}
	t_1 = i \int_0^{\pi/2} \frac{d\theta}{\sqrt{1-(1-k^2)\sin^2 \theta}}.
\end{equation}

This is exactly the definition of the complementary elliptic integral:

\begin{equation}
	t_1 = i K'.
\end{equation}

With the universal period $T = 4 K$ and the first pole at $t_1 = i K'$ established, locating another pole $t_2$ is an immediate consequence of the periodicity of $\dot{\phi}(t)$. Because the dynamics at $t = 2 K$ are the exact mirror of $t=0$ with identical angular speed, the integration path from this point to the complex infinity is mathematically identical to the first pole’s derivation. The travel time required to reach the singularity is again purely imaginary: $i K'$.

Therefore, shifted by the real half-period, the second pole lies universally at:

\begin{equation}
	t_2 = 2 K + i K'.
\end{equation}

Figure \ref{fig_contour} shows the location of the poles $t_1$ and $t_2$ of the two regimes. The universal period $T = 4 K$ allows for symmetric locations of the singularities and the residues of both regimes.

Having evaluated the interval $i K'$ for the pendulum to traverse directly into imaginary time, we use this to establish another symmetrical property of the angular velocity. Starting from the initial condition $\phi_0 = 0$, the pendulum in real time can evolve its trajectory $\phi(t)$ and exhibits swinging/stopping/spinning depending on the initial speed $\omega_0$. However, in imaginary time there is only the singularity $\phi \to \pm i \infty$ that the trajectory is going to reach. Starting from $\phi_0 = 0$, it takes imaginary time $i K'$ to reach the pole, another $i K'$ back to zero, another $i K'$ to swing to the negative side, and finally another $i K'$ back to complete the cycle. In other words, the imaginary period is $T = 4 i K'$, and there is only a \textquote{swinging} pendulum in \emph{imaginary time}.

Because the dynamics are always \textquote{swinging}, conservation of energy dictates that the angular velocity comes back to the same magnitude after a half-swing, analogous to Eq.~(\ref{eq:halfperiod_swing}). Therefore, $\dot{\phi}$ is anti-periodic over a half-period in imaginary time:

\begin{equation}
 \dot{\phi}(t) = - \dot{\phi}(t +  2 i K').\label{eq:halfperiod_imaginary}
\end{equation}

\subsection{The Residues}

A residue is the coefficient of the $\frac{1}{t-t_0}$ term in the Laurent series of a function $f(t)$ around an isolated singularity $t_0$. If $f(t) = g(t)h(t)$ where $h(t)$ is smooth and non-zero at $t_0$, the product rule gives:

\begin{equation}
	\text{Res}(g \cdot h, t_0) = \left[\text{Res}(g,t_0)\right] h(t_0).
\end{equation}

Therefore, we find the residues of $\omega(t) = \dot{\phi}(t)$ first, and the term $\frac{1}{T} e^{- i n \Omega_0 t}$, which is smooth and non-zero, can be multiplied to its residues afterward.

Instead of relying on explicit solutions to find the nature of these singularities, we ask the differential equation itself. For a singularity at $t_1$ with a simple pole:

\begin{equation}
	\dot{\phi} \sim \frac{A}{t - t_1}
\end{equation}

Differentiating gives $\ddot{\phi} = - \frac{A}{(t-t_1)^2}$. Squaring gives $\frac{1}{2} \dot{\phi}^2 \sim \frac{A^2}{2(t-t_1)^2}$. Substituting this dominant term into the energy equation yields $\cos \phi \approx \frac{A^2}{2 (t-t_1)^2}$. With the identity $\sin \phi = \sqrt{1 - \cos^2 \phi}$, we may neglect the constant $1$ because $\cos \phi$ is approaching singularity. This means $\sin \phi \approx i \frac{A^2}{2(t - t_1)^2}$.

Substituting these back into the original ODE $\ddot{\phi} = - \sin \phi$:

\begin{equation}
	- \frac{A}{(t-t_1)^2} = - i \frac{A^2}{2(t-t_1)^2} \Longrightarrow A = - 2 i.
\end{equation}

With the residue of $\dot{\phi}(t)$ at the pole $t_1$ established, the residue at $t_2 = t_1 + 2 K$ is only a matter of symmetric or anti-symmetric nature of $\dot{\phi}(t)$. The half-period shift $2 K = T/2$ from $t_1 \to t_2$ allows us to apply the half-period symmetry in Eq.~(\ref{eq:halfperiod_swing})--(\ref{eq:halfperiod_spin}). In other words, $\dot{\phi}$ only changes or preserves its sign. Because the residues are the limiting behavior of the function $f(t)$ itself, if it changes sign, so does its residue.

Therefore, the residues at the two poles of the function $f(t) = \frac{1}{T} \dot{\phi}(t) e^{- i n \Omega_0 t}$ are:

\begin{equation}
\begin{aligned}
\text{swinging: } \text{Res}_1 = - \frac{1}{T} 2 i e^{-i n \Omega_0 (i K')}, & \text{ Res}_2 = +\frac{1}{T} 2 i e^{- i n \Omega_0 (2 K + i K')}, \\
 \text{spinning: } \text{Res}_1 = - \frac{1}{T} 2 i e^{-i n \Omega_0 (i K')}, & \text{ Res}_2 = -\frac{1}{T} 2 i e^{- i n \Omega_0 (2 K + i K')}.
\end{aligned}
\end{equation}

Note the exponential $\exp(- i n \Omega_0 2 K)$ in the second residue. Substituting $\Omega_0 = 2 \pi/(4 K)$, the exponential becomes $\exp(- i n \pi) = (-1)^n$. This sign alternating term forces the sum of the residues to vanish at a certain parity. Namely,

\begin{equation}
	\begin{aligned}
		\text{swinging: } & \sum \text{Res} = - \frac{1}{T} 4 i e^{n \Omega_0 K'}, \; n \in \text{odd}, \\
		\text{spinning: } & \sum \text{Res} = - \frac{1}{T} 4 i e^{n \Omega_0 K'}, \; n \in \text{even}.
	\end{aligned}
\end{equation}

\subsection{The Complex Contour}

As shown in Fig.~\ref{fig_contour}, we choose the rectangular contour to enclose the strip where we know the poles reside: i) Real boundaries: $0 \le t \le T$, ii) Imaginary boundaries: $0 \le t \le 2 i K'$.

The contour consists of four straight segments. Let us traverse them counter-clockwise:

\begin{enumerate}
	\item Bottom Edge: We evaluate the integral $I_1 = \int_0^T f(t) dt$. By definition, this is exactly $w_n$. Hence, $I_1 = w_n$.
	\item Right Edge: Moving up the imaginary direction, we denote this integral $I_2$.
	\item Top Edge: We traverse backward in the real direction. We need to evaluate the integral $
	I_3 = \int_T^0 f(t) dt$. Along this path, $t = x + 2 i K'$, where $x$ goes from $T \to 0$. Therefore, by change of variable $t \to x$:
	
	\begin{equation}
	I_3 = \int_T^0 \frac{1}{T} \dot{\phi}(x + 2 i K') e^{- i n \Omega_0 (x + 2 i K')} dx.
	\end{equation}
	
	The technique is to express $I_3$ in terms of $w_n$. To do so, we use the imaginary anti-symmetry in Eq.~(\ref{eq:halfperiod_imaginary}), $\dot{\phi}(x + 2 i K') = - \dot{\phi}(x)$. Therefore,
	
   \begin{equation}
   I_3  =  - \int_T^0 \frac{1}{T} \omega(x) e^{-i n \Omega_0 (x + 2 i K')} dx  = e^{2 n \Omega_0 K'} \int_0^T \frac{1}{T} \omega(x) e^{-i n \Omega_0 x} dx.  
   \end{equation}
	
where we flip the bounds of the integration to absorb the negative signs and factor out the $x$-independent exponential. The integral above is precisely $w_n$ with just the variable labeled $x$ instead of $t$. Therefore,

	\begin{equation}
	I_3 = e^{2 n \Omega_0 K'} w_n.	
	\end{equation}

 \item Left Edge: We move down the imaginary direction. We denote this integral $I_4$.

\end{enumerate}

Combining the integrals from the edges together, the $I_2$ and $I_4$ exactly cancel, leaving only the surviving contributions:

\begin{equation}
		 I = I_1 + I_3  = w_n (1 + e^{2 n \Omega_0 K'})  = w_n 2 \cosh(n \Omega_0 K') e^{n \Omega_0 K'}.
\end{equation}

\subsection{The Cauchy Residue Theorem}

With the residues and the contour integral made explicit, we apply the Cauchy residue theorem in Eq.~(\ref{eq:Cauchy}) to evaluate the exact Fourier coefficient for both regimes:

\begin{equation}
\begin{aligned}
\text{swinging:} & \quad 2 w_n \cosh( n \Omega_0 K') e^{n \Omega_0 K'} = 2 \pi i \left( - \frac{1}{T} 4 i  e^{n \Omega_0 K'}\right), \; n \in \text{odd}, \\
 \text{spinning:} & \quad 2 w_n \cosh( n \Omega_0 K') e^{n \Omega_0 K'} = 2 \pi i \left( - \frac{1}{T} 4 i  e^{n \Omega_0 K'}\right), \; n \in \text{even}.
\end{aligned}
\end{equation}

Solving for $w_n$ and substituting $T \to 2 \pi/\Omega_0, \; K' \to \kappa$ yields:

\begin{equation}
	\begin{aligned}
		\text{swinging:} & \quad w_n = \frac{2 \Omega_0}{\cosh(\kappa n \Omega_0)}, \; n \in \text{odd}, \\
		\text{spinning:} & \quad w_n = \frac{2 \Omega_0}{\cosh(\kappa n \Omega_0)}, \; n \in \text{even}.
	\end{aligned}
\end{equation}

Because of the symmetry $w_{+n} = w_{-n}$, the complex Fourier series simplifies to cosine Fourier series. We seek the Fourier cosine transform of the separatrix (stopping) regime as well. This can be done by setting $k = 1, E = 2k^2 - 1$ in the energy integral:

\begin{equation}
\quad  dt = \frac{d\phi}{\sqrt{2 + 2 \cos \phi}}, \quad k \to 1.
\end{equation}

Integrating and applying the initial condition $\phi_0 = 0$ gives the solution $\phi(t) = 2 \arcsin[\tanh(t)]$. Its derivative yields $\omega(t) = 2/\cosh(t)$, which has the standard Fourier transform:

\begin{equation}
	\text{stopping:} \quad \omega(t) = \int_{-\infty}^{+\infty} \frac{e^{i \Omega t} d\Omega}{\cosh(\kappa \Omega)}.
\end{equation}

Here, $\kappa \to \pi/2$ as $k \to 1$ is a standard property of the complementary elliptic integral. Rescaling the time by substituting $T \to 4 K/\Omega_L$ and $\kappa \to \kappa/\Omega_L$ thereby reintroduces the linear oscillation parameter $\Omega_L$. Finally, integrating the velocity with respect to time gives the angular displacement $\phi(t)$ in terms of the exact Fourier series.

\section{Result and Discussion}

Following Method, an expression that gives the exact time and frequency domain solution across all regimes is:

\begin{equation}
	\omega(t)  = \sum_{\mathcal{P}}  \frac{ 2  \Omega_0   e^{i n \Omega_0 t}  }{\cosh(\kappa n \Omega_0/\Omega_L)}   , \quad \mathcal{P}  = \begin{cases}
	n \in \text{odd} & \text{swinging } \; k < 1  \\
	n \in \text{even} & \text{spinning } \;  k > 1 \\
	\text{continuous} & \text{stopping } \; k = 1
\end{cases}. \label{eq:omega_exact}
\end{equation}

The parameters $\kappa$, fundamental frequency $\Omega_0$, and period $T$ are defined  as follows:

\begin{equation}
	\kappa \equiv K(\sqrt{1-k^2}), \quad \Omega_0 = \frac{2\pi}{T}, \quad T \equiv \frac{4 \text{Re}[K(k)]}{\Omega_L}.\label{eq_kappa_T}
\end{equation}

Here, $K(k)$ is the complete elliptic integral of the first kind~\cite{NIST:DLMF,Chachiyo2026}.  Formally, $\kappa$ and $\Omega_0$ remain well-defined positive real quantities for the full range of modulus $k$, so that the definitions above apply to all classes of motion. The angular displacement $\phi(t)$ can be obtained via a simple integration and simplified into sine Fourier series or transform (see Appendix \ref{appendix:phi}).

The Fourier series in Eq.~(\ref{eq:omega_exact}) is valid for the stopping regime where $k \to 1$ as well because $\Omega_0 \to 0$ in this limit, turning its discrete summation into a Riemann sum (e.g., an integration), or equivalently the Fourier series into a Fourier transform. This can be written explicitly as follows.

\begin{equation}
	 \omega(t) = \int_{-\infty}^{+\infty} \frac{e^{i \Omega t} d\Omega}{\cosh(\kappa \Omega/\Omega_L)}, \qquad \text{stopping }\; k = 1,  \; \kappa = \pi/2.
\end{equation}

We discover that all regimes arise from a single spectral kernel, with parity selection distinguishing the periodic motions and the separatrix representing their discrete-to-continuum limit. Regime changes thus correspond to symmetry-driven reorganizations in frequency space rather than changes in the underlying spectral structure, with the stopping trajectory representing the continuum limit.

An early appearance of a $1/\cosh(\cdots)$ spectral factor was reported by Chirikov~\cite[$\S2.1$,Eqs.(2.4),~(2.6)]{Chirikov1979} where the author relied on the Jacobi elliptic framework to consider solutions of $\phi(t)$ for the swinging and spinning regimes. It was found that their Fourier series can be made as similar to each other as possible by introducing half the frequency of the mean rotation. The near-separatrix solution~\cite[$\S2.4$, Eq.~(2.32)]{Chirikov1979} was then approximated as the Fourier series of both regimes (e.g., made similar previously). However, these results were not explicitly formulated as a single unifying spectral-kernel description encompassing the swinging, spinning, and exact-separatrix (stopping) regimes.

By completely bypassing the traditional Jacobi elliptic functions, and relying on the symmetrical structure of the complex time plane, our derivation shows that the symmetrical spectral structure is not merely an algebraic artifacts of elliptic integrals, but fundamentally connected to the dynamical symmetries of the system.

As a pedagogical example of how looking at a physics problem in different ways can eventually lead to its solution, we summarize the guiding principles. The unified frequency-domain structure becomes transparent only after three key perspective shifts.

\begin{enumerate}	
	\item \textit{Choice of initial condition.} 
	Instead of adopting the traditional formulation in which the pendulum is released from rest at an amplitude $\phi_0$, we consider the motion initiated at the bottom. This perspective naturally promotes the initial angular speed $\omega_m$ to a characteristic parameter, enabling continuous transitions between all three classes of motion.	
	
	\item \textit{Choice of dynamical variable.} 
	The challenge in finding the exact Fourier coefficients arises from the fact that the Jacobi elliptic function is embedded within the $\arcsin$ function. However, we have found that the $\arcsin[\cdots]$ dependence of $\phi(t)$ can be avoided if one initially focuses on  the  angular velocity $\omega(t)$. Once its spectral content is obtained, one gains access to the frequency-domain solution of $\phi(t)$ via simple integration.
	
	\item \textit{Choice of period in the spinning regime.} 
	The motion can be decomposed exactly into a linear phase drift plus a periodic phase oscillation. The oscillatory component remains strictly periodic. Exploiting this periodicity, we adopt twice the primitive oscillation period as the fundamental period—an admissible spectral choice—which aligns the harmonic lattice of the spinning solution with that of the swinging solution. This step is not merely technical: it reveals that both regimes share an identical analytic coefficient structure and differ only by parity selection. It also permits the separatrix to be identified as the continuum limit common to both regimes.	
	
\end{enumerate}

Together, each perspective shift reveals the solution layer by layer—(i) disentangling regime classification, (ii) removing the $\arcsin$ obstruction, and (iii) exposing parity and the continuum limit—thereby unifying the exact spectral structure across all regimes of motion.

\begin{figure*}
	\begin{center}
		\includegraphics[width=\textwidth]{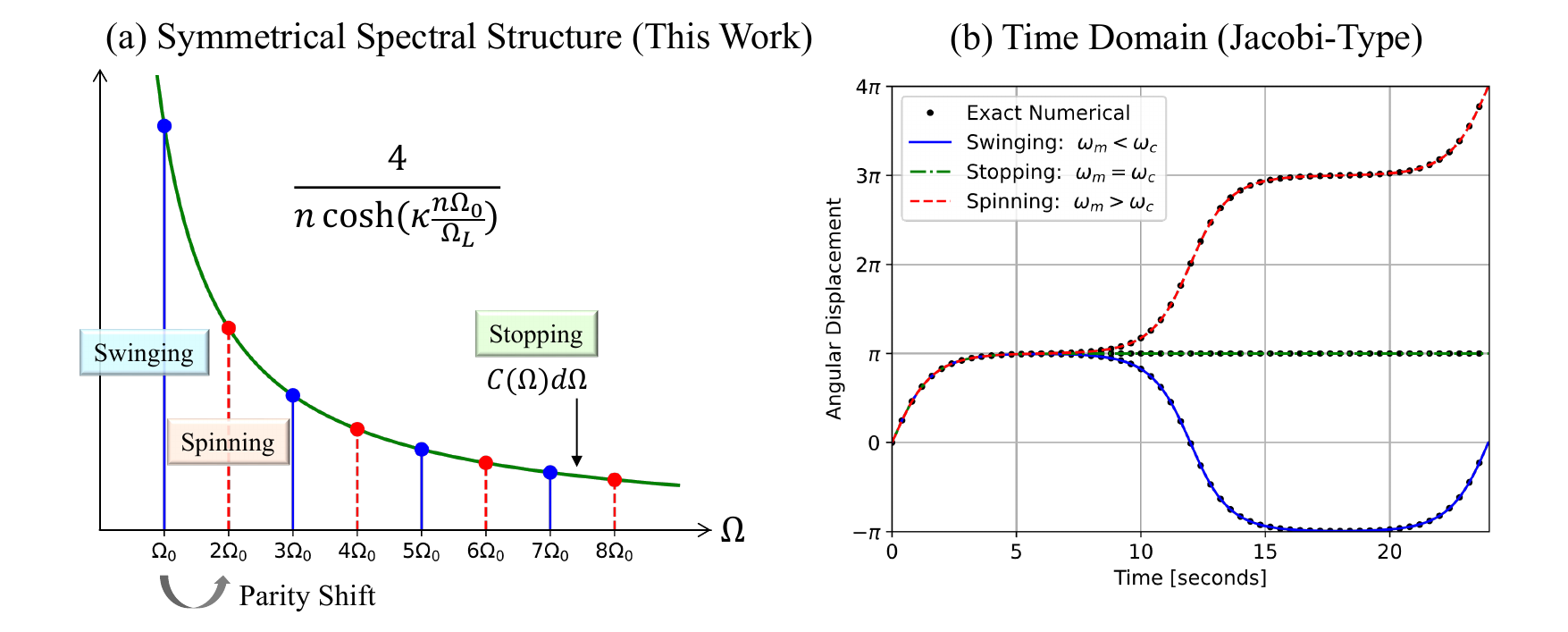}
		\caption{\label{fig_3S} 
			\textbf{Frequency- and time-domain representations of pendulum-like dynamics.} 
			(a) Symmetrical frequency-domain representation, showing that all three regimes share the same analytic spectral structure, while differing only by parity selection of harmonics (odd for swinging, even for spinning) and by a discrete-to-continuum transition at the separatrix.
			(b) Time-domain characteristics of the three regimes, comparing the spectral solutions in Eqs.~(\ref{eq_swinging})--(\ref{eq_spinning}) with exact numerical results obtained from Jacobi elliptic functions.
		}
	\end{center}
\end{figure*}

\section{Validation}

Figure~\ref{fig_3S}(b) shows examples of $\phi(t)$ from this work. As clearly seen, the pendulum starts at the bottom and continues to rise, but eventually branches off into three distinct patterns. Quantitatively, the exact numerical values obtained using solutions in the form of Jacobi elliptic functions are also plotted in Fig.~\ref{fig_3S}(b). The agreement is found to be within 14 significant digits, a typical limit of computer precision. Therefore, the solutions in this work are also exact. 

Traditionally, the perturbation method has been the primary theoretical tool for studying the frequency content of the nonlinear pendulum \cite{Fulcher1976, Zilio1982}. However, its scope has been limited to a swinging motion where the pendulum is initially at rest, positioned at an amplitude $\phi_0$. As a result, the $\phi(t)$ is often written as a Fourier series of odd harmonics in perturbation methods. For example, a recent experimental study \cite{Hinrichsen2020} has used the following coefficients:

\begin{equation}
	\begin{aligned}
		A_1 & = +\left( \phi_0 + \frac{1}{192}\phi_0^3 + \frac{17}{120 \cdot 512} \phi_0^5 + \cdots \right) \\
		A_3 & = - \left( \frac{1}{192} \phi_0^3 + \frac{1}{6 \cdot 512} \phi_0^5 + \cdots \right) \\
		A_5 & = +\left( \frac{1}{40 \cdot 512} \phi_0^5 + \cdots \right).
	\end{aligned}
\end{equation}

\begin{figure}[t]
	\begin{center}
		\includegraphics[width=0.8\textwidth]{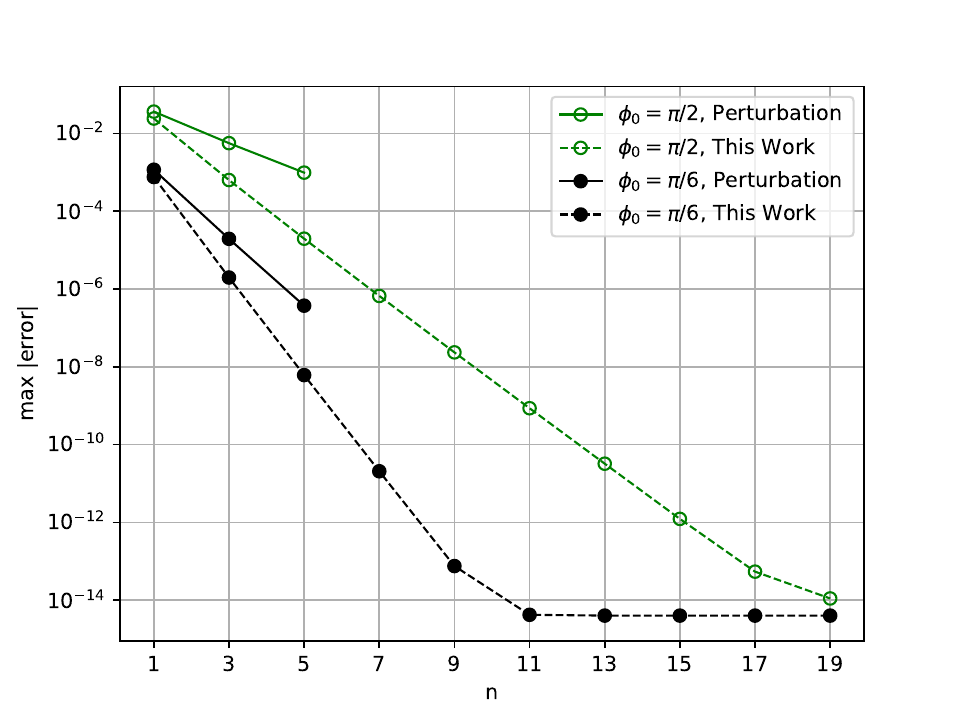}		
		\caption{\label{fig_maxerr}
			\textbf{Convergence of frequency-domain solutions.}
			Maximum error of the angular displacement $\phi(t)$ relative to exact numerical solutions, comparing the perturbation method with the exact spectral solution presented in this work.
			Results are shown for two initial amplitudes: $\phi_0 = \pi/6$ (solid circles) and $\phi_0 = \pi/2$ (open circles).
			The horizontal axis denotes the perturbation order $n$ for the perturbative approach, or equivalently the number of harmonics retained in the spectral solution.
			Both methods exhibit exponential convergence, but the closed-form coefficients in Eq.~(\ref{eq_swing_spin_coef}) achieve machine precision with fewer harmonics, particularly at larger amplitudes.
		}		
	\end{center}
\end{figure}

Here, $\{A_1, A_3, A_5\}$ are the coefficients of the odd harmonics. They are power series in the amplitude, $\phi_0$, and are shown up to order 5. In general, higher orders are possible but, as previously noted, the expression would be increasingly more complex.

More broadly, the perturbation method offers a versatile mathematical structure that is applicable beyond the energy-conserving pendulum in this study. Therefore, the method is still an invaluable theoretical tool for a wide range of nonlinear dynamics. However, it is instructive to compare the performance of the frequency content given by the perturbation method and that of Eq.~(\ref{eq_swing_spin_coef}) in this work.

Figure~\ref{fig_maxerr} shows the maximum error of $\phi(t)$ from the perturbation method relative to the exact numerical solutions. Two amplitude cases are presented: small amplitude $\phi_0 = \pi/6$ (solid circle) and large amplitude $\phi_0 = \pi/2$ (open circle). The horizontal axis indicates the order of the perturbation, with available values $n = 1, 3, 5$. The graph also shows the maximum error of $\phi(t)$ in this work as a function of the number of included harmonics. Together, these graphs provide a comparative view of the convergence performance between the perturbation method and the solution presented in this work.

Both methods exhibit rapid exponential convergence, as indicated by the straight line on the logarithmic scale. Exponential convergence is a desirable feature, particularly given that many periodic functions---such as the square wave---exhibit only polynomial convergence. As expected, better performance is observed at smaller amplitudes, as indicated by the steeper decline in the error. However, the swinging solution in Eq.~(\ref{eq_swinging}) appears to converge faster as the number of harmonics increases. For the amplitude $\phi_0 = \pi/6$, it takes 11 harmonics for the error to reach machine precision. Larger amplitudes require more harmonics, but the coefficients $c_n$ are already given in closed form and can always be evaluated to the desired precision.

Taken together, the results presented here establish a unified, exact, and frequency-domain description of all regimes of the nonlinear pendulum---swinging, stopping, and spinning. The derived expressions reproduce the time-domain solutions to within machine precision while offering direct access to the spectral structure of the motion through simple elementary functions. Their convergence properties outperform traditional perturbation approaches, which require increasingly complex higher-order expansions to achieve comparable accuracy.

\section{Pedagogical Overview of Quantum Analogues}\label{sec:quantum_analogue}

Here, we give pedagogical overviews of the state-of-the-art of the pendulum-like systems, specifically in the context of the spectral solution presented in this work.

Beyond classical physics, non-intuitive applications of the nonlinear pendulum are found in emerging fields of quantum physics, such as the superconducting Josephson junction \cite{MacDonald1983, Mangin2016} and, most recently, Bose--Einstein condensate tunneling \cite{Marino1999,Pigneur2018} as shown in Fig.~\ref{fig_pendulum-like}.

For example, a Josephson junction (JJ) is a device consisting of two superconducting islands separated by a thin barrier, which can be a normal metal, a narrow superconducting wire, or a thin insulator \cite{Rasmussen2021}. Unlike a capacitor (C) or an inductor (L), which are linear electronic devices, the Josephson junction is nonlinear. JJs have  been used extensively to construct qubits \cite{Shnirman1997,Rasmussen2021, Kim2024}, the fundamental units of quantum computers, in devices like IBM's quantum processors \cite{AbuGhanem2025}, Google's Sycamore Processor \cite{Arute2019}, and Zuchongzhi processors \cite{Gao2025}.

The dynamics of a Josephson junction can be understood via BCS theory \cite{Bardeen1957}, in which Cooper pairs exist on both superconducting islands. The number of pairs on each island need not be equal, leading to an imbalance of quantum particles between the two sides of the junction. This imbalance can change over time through the tunneling process. Since Cooper pairs are charged particles, their tunneling generates an electric current; likewise, the imbalance gives rise to an electric voltage.

An important quantum mechanical parameter is the phase difference $\phi$ between the wavefunctions of the two superconducting islands. This is directly analogous to the angle $\phi$ in the classical pendulum. Following standard circuit theory, the dynamical equation for an ideal Josephson junction (JJ) can be written as \cite{Blackburn2016,Roth2023}:

\begin{equation}
	\frac{d^2}{dt^2} \phi(t) + \frac{2e I_c}{\hbar C} \sin(\phi) = 0.
\end{equation}

Here, $C$ is the stray capacitance due to the junction’s geometrical arrangement, and $I_c$ is the critical tunneling current, which depends on the material, thickness, and area of the junction. As the phase difference $\phi$ fluctuates—analogous to the dynamics of the angle $\phi$ in the pendulum—so do the current through and the voltage across the junction. The current and the voltage are related to wavefunctions phase difference as follows.

\begin{equation}
	I = I_c \sin \phi,\quad V = \frac{\hbar}{2 e} \frac{d\phi}{dt}.
\end{equation}

In particular, the voltage across the junction plays similar role as the angular velocity of the nonlinear pendulum. In  the context of the formalism presented in this work, the dynamics of the ideal Josephson junction can be deduced as follows.

Consider the initial condition where the phase difference is $\phi = 0$, and the dynamics are initiated solely by an imbalance of negatively charged Cooper pairs between the two islands. This initial imbalance can also be interpreted as an initial voltage $V_m$ across the junction. The value of $V_m$ determines the regime of the dynamics. If $V_m$ is too low, the voltage simply fluctuates around a zero average. This behavior is analogous to the swinging motion of a pendulum, where the angular velocity $\omega(t)$ oscillates back and forth with zero mean.

There exists a critical voltage $V_c$ that separates the dynamics into three regimes, analogous to the critical angular velocity $\omega_c$ defined earlier. If $V_m = V_c$, the voltage no longer fluctuates but instead decays toward zero. This corresponds to the separatrix (stopping) regime, where the angular velocity $\omega(t)$ also decays to zero. If $V_m > V_c$, voltage fluctuations resume, but with a non-zero average. This corresponds to the spinning motion of the pendulum, where $\omega(t)$ oscillates with a non-zero mean. The average voltage across the junction that persists over time can be precisely determined, analogous to the mean angular velocity $\bar\omega = 2 \Omega_0$ in the spinning solution.

In summary, a Josephson junction is a superconducting device that exhibits a time-varying voltage across the junction, analogous to the changing angular velocity of a pendulum. The behavior of this voltage can be categorized into three distinct regimes: fluctuations around a zero average, continuous decay toward zero, and fluctuations with a persistent non-zero average.

Another example of a quantum mechanical analogue involves two coupled Bose--Einstein condensates (BECs) of neutral bosonic atoms trapped in a double-well potential, known as a bosonic Josephson junction (BJJ) \cite{Smerzi1997,Milburn1997,Albiez2005,LeBlanc2011}. The bosons can tunnel between the wells, and the key observable variables are the atom number imbalance and the relative phase between the two condensates. The dynamics of the BJJ, modeled as a pendulum, have been extensively studied in the notable work of Pigneur and Schmiedmayer \cite{Pigneur2018}. Their analysis will be briefly revisited here to highlight its similarity to the JJ system.

Unlike Cooper pairs in JJ, the bosons in BJJ are neutral; thus, the imbalance between the two sites does not give rise to the electric voltage. Therefore, the normalized atom number imbalance $n$, and the phase difference $\phi$ are the key conjugating variables describing the dynamics.

\begin{equation}
	n = \frac{N_L - N_R}{N_L + N_R}, \quad \phi = \phi_L - \phi_R.
\end{equation}

When the imbalance is appreciably small, $n(t) \ll 1$, the dynamical equation can be written as:

\begin{equation}
	\frac{d^2}{dt^2} \phi(t) + [\frac{2 J}{\hbar}\sqrt{\Lambda + \lambda}]^2 \sin(\phi) = 0.
\end{equation} 

Here, $\{J, \Lambda, \lambda\}$ are physical constants characterizing tunneling energy and inter-atomic interaction. Similar to JJ, where the voltage plays the role of angular velocity in classical pendulum, $n(t)$ in BJJ is related to the phase difference as follows.

\begin{equation}
	n(t) = \frac{\hbar}{2 J} \frac{1}{\Lambda + \lambda} \frac{d\phi}{dt}.
\end{equation} 

Therefore, $n(t)$ is analogous to $\omega(t)$ in classical pendulum. As expected, the imbalance $n(t)$ can vary over time in three distinctive regimes:  fluctuations around a zero average, continuous decay toward zero, and fluctuations with a persistent non-zero average. 

Having discussed the specific examples of the JJ and BJJ, we can extract general features that give rise to pendulum-like dynamics in quantum systems. First, the system consists of two weakly coupled quantum compartments, with the phase difference \(\phi\) between the wavefunctions in each compartment serving as the key dynamical variable. Second, the weak coupling permits quantum particles to tunnel between compartments, leading to a time-dependent imbalance in particle number. Third, it is energetically favorable for the system to remain in phase; that is, the potential energy is minimized when \(\phi = 0\). These three features may give rise to pendulum-like dynamics, as exemplified by the JJ and BJJ systems\textemdash cases in which the solutions presented in this work offer a unified and analytically tractable framework for their description.

\section{Conclusion and Outlook}

We have presented an exact time- and frequency-domain solution of energy-conserving pendulum-like dynamics that unifies swinging, stopping, and spinning motions within a highly symmetrical spectral structure. By exposing the spectral kernel common to all regimes via complex time and the Cauchy residue theorem, this approach reveals that qualitative changes in motion arise not from altered frequency scales or deformed spectra, but from systematic reorganizations of spectral weight—through parity selection and, at the separatrix, through a dynamical collapse of the fundamental frequency.

Beyond the examples presented here, this spectral unification can be expanded to cover other branches of physics and engineering applications. In astrophysics, the same symmetrical spectral structure may appear in the so-called neutrino flavor pendulum that governs collective neutrino oscillations in dense environments such as core-collapse supernovae and neutron-star mergers. In engineering and classical mechanics, related separatrix structures arise in systems such as the Duffing oscillator and in rigid-body dynamics through the Dzhanibekov effect (or tennis-racket theorem)~\cite{Chachiyo2026taxonomy}. By focusing on the spectral properties of the separatrix, the present approach offers a complementary alternative to traditional treatments based on Jacobi elliptic functions. More importantly, it suggests that complex nonlinear motions in these systems may be analyzed and potentially controlled directly through their frequency-domain signatures.

Perhaps more intriguingly, the spectral interpretation developed here also invites a different perspective on the emergence of chaos, closely related to Chirikov's analysis of separatrix splitting under periodic perturbations~\cite{Chirikov1979}. In the present framework, the separatrix appears as the continuum limit in which discrete harmonic structure dissolves and all frequencies become accessible. From this viewpoint, chaotic motion need not be regarded solely as irregular behavior in the time domain, but rather as a dynamical sampling of an underlying spectral structure in the frequency domain. When weak damping or external driving is introduced, the system may effectively become trapped near this spectral boundary, attempting to realize both swinging and spinning dynamics simultaneously. The apparent complexity of the motion may then arise from interference between these competing spectral organizations. Extending this idea further, systems with \emph{multiple interacting separatrices} may generate overlapping spectral continua, producing dynamics that resemble hyper-chaos or even turbulence. In this sense, the separatrix may represent not only the boundary between simple regimes of motion, but also a spectral gateway through which ordered nonlinear dynamics give way to chaos.

\section*{Acknowledgments}
The author thanks the Potjanachaisak family for their hospitality during a summer research visit in Chiang Khruea, Sakon Nakhon Province, Thailand. Discussions with Dr. Hathaithip Chachiyo---whose non‑specialist questions prompted a return to fundamentals---ultimately led to the perspective shifts that enabled the exact spectral solutions in this work.

\section*{Author Declarations}
The author has no conflicts to disclose.

\section*{Data Availability}
Python validation and teaching guide are publicly available at https://github.com/teepanis/nonlinear-pendulum.

\appendix

\section{Angular Displacement $\phi(t)$}\label{appendix:phi}

\begin{align}
	\text{swinging: }  \quad &  \phi(t) = \sum_{n \in \text{odd}}  c_n \sin( n \Omega_0 t), \label{eq_swinging} \\
	\text{stopping: }   \quad &   \phi(t)  = \int_0^\infty \!\!\! d\Omega \, C(\Omega) \sin (\Omega t), \label{eq_stopping_freq} \\
	\text{spinning: } \quad &  \phi(t)  =  2 \Omega_0 t +  \sum_{n \in \text{even}} c_n \sin(n \Omega_0 t ) . \label{eq_spinning} 
\end{align}

The summations exclude $n=0$. The spectral coefficients $\{c_n,  C(\Omega)\}$ are:

\begin{align}
	\text{swinging, spinning: }  \quad &  \quad\; c_n  =  \frac{4}{n \cosh(\kappa n \Omega_0/\Omega_L)}, \label{eq_swing_spin_coef} \\
	\text{stopping: } \quad &  C(\Omega)  =   \frac{2}{\Omega \cosh(\kappa \Omega/\Omega_L)}. \label{eq_stopping_coef}
\end{align}

\bibliographystyle{vancouver}  
\bibliography{symmetrical_spectral_v1da.bib}


\end{document}